\documentclass[11pt,a4paper]{article}

\usepackage[margin=2.5cm]{geometry}
\usepackage[numbers]{natbib}
\usepackage{amsmath}
\usepackage{amssymb}
\usepackage{amsthm}
\usepackage{bm}
\usepackage{graphicx}
\usepackage{caption}
\usepackage{multirow}
\usepackage{authblk}
\usepackage[colorlinks=true, allcolors=blue]{hyperref}
\theoremstyle{plain}

\theoremstyle{definition}

\newtheorem{example}{Example}

\begin{document}

\title{Quantum AS-DeepOnet: Quantum Attentive Stacked DeepONet for Solving 2D Evolution Equations}

\author[1]{Hongquan Wang}
\author[2,$\dagger$]{Hanshu Chen}
\author[3]{Ilia Marchevsky}
\author[1,2,$\dagger$]{Zhuojia Fu}

\affil[1]{School of Mathematics, Hohai University, Nanjing, Jiangsu 211100, China}
\affil[2]{College of Mechanics and Engineering Science, Hohai University, Nanjing, Jiangsu 211100, China}
\affil[3]{Department of Applied Mathematics, Bauman Moscow State Technical University, Moscow 105005, Russia}
\affil[$\dagger$]{Corresponding authors: \texttt{chenhanshu@163.com} (H.\ Chen); \texttt{paul212063@hhu.edu.cn} (Z.\ Fu)}

\date{}

\maketitle

\begin{abstract}
DeepONet enables retraining-free inference across varying initial conditions or source terms at the cost of high computational requirements. This paper proposes a hybrid quantum operator network (Quantum AS-DeepOnet) suitable for solving 2D evolution equations. By combining Parameterized Quantum Circuits and cross-subnet attention methods, we can solve 2D evolution equations using only 60\% of the trainable parameters while maintaining accuracy and convergence comparable to the classical DeepONet method.
\end{abstract}


\section{Introduction}
Evolution equations are fundamental models for capturing temporal dynamics in science and engineering. Classical numerical methods such as finite element methods are reliable under sufficiently fine spatial-temporal discretization, but can become computationally and memory intensive in multiscale, complex-geometry, and long-time integration settings.
Recently, machine learning has been used for fast approximation and surrogate modeling of PDEs. A representative approach is physics-informed neural networks \cite{2019Raissi, 2020Pang}, which encode governing equations, boundary conditions, and initial conditions in the training objective. This approach reduces the need for labeled data, but optimization can be difficult and sensitive to stiffness and loss imbalance. Another representative approach is DeepONet \cite{2021lulu}, which learns nonlinear operators between function spaces via a branch--trunk decomposition and has demonstrated strong generalization on a variety of PDE benchmarks. However, when an input function must be discretized using many sensors or when the input dimension increases, DeepONet can require substantially more parameters and computational cost, and may suffer from the curse of dimensionality.

Motivated by the superposition principle and entanglement in quantum Mechanics, quantum methods can implement high-dimensional feature maps in exponentially large Hilbert spaces and provide expressive presentations\cite{2019Schuld,2019V}. Quantum approaches have also been explored for a range of learning and inference tasks, including quantum reinforcement learning \cite{2021Kwak}, quantum Boltzmann machines \cite{2018Amin}, quantum principal component analysis \cite{2021Xin} and quantum support vector machines \cite{2024li}. In machine learning settings, quantum models are realized by Parameterized Quantum Circuits (PQCs) \cite{2025Farea, 2024Baker, 2019Benedetti}. The integration of PQCs into DeepONet utilizes quantum Hilbert spaces to surpass the accuracy and parameter efficiency of classical models. This synergy establishes a more scalable approach for approximating 2D evolution equations.

However, studies on quantum DeepONet are still limited to low-dimensional inputs, with their application to 2D evolution equations remaining largely unexplored. Wang et al. \cite{2025Wang} proposed QuanOnet with trainable frequencies. It improves robustness for high-frequency or irregular cases. But when the frequency range of the problem is wide, it needs deeper quantum layers to cover all frequencies. In addition, when the input to the quantum branch network grows, the circuit depth must increase to encode all inputs because of the repeated encoding technique. Xiao et al. \cite{2025Xiao} proposed Quantum DeepONet with unary encoding. It reduces the evaluation cost. However, for 2D evolution equations, the branch net still needs an affine mapping to map the high-dimensional input to the number of qubits. In the NISQ era, the available qubits are usually far fewer than the input dimension of  2D evolution equations. These problems need high-resolution inputs to capture high frequencies or small-scale dynamics. Unary encoding also increases the required number of qubits. This limits wide networks on current quantum hardware.

To address high-dimensional operator learning in 2D evolution equations, we propose Quantum Attentive Stacked DeepONet. It is inspired by stacked DeepONet. Stacked DeepONet was built to better match operator approximation theory. Later numerical studies showed that it needs more trainable parameters. We find that this stacked structure can be adapted to parameterized quantum circuits. In the branch net, we stack several hybrid quantum layers to learn the full high-dimensional input in blocks. We then use adapted Efficient Channel Attention (ECA) to capture cross-subnet relations among tokens from different quantum sub-networks with very few trainable parameters. It learns weights for these tokens and enables global coordination of local features. In the trunk net, we use the same hybrid quantum layer as the core operator. It maps low-dimensional space-time inputs (e.g., $(x, y, t)$) to the same basis function space as the branch net. Finally, we take an element-wise inner product with the branch features. This hybrid classical--quantum model can accurately fit the PDE operator mapping. This paper is organized as follows: In Section \ref{architecture}, the architecture of Quantum AS-DeepOnet is given. The numerical experiments of the method are shown in Section \ref{numerical experiments}.

\section{Methodology}\label{architecture}






Let $\mathcal{A}(j,k)$ denote the set of affine mappings from $\mathbb{R}^j$ to $\mathbb{R}^k$. Define the space $\mathcal{S}_{m,n}^\sigma$ as the affine composition space induced by the non-linear activation function $\sigma$:
\begin{equation*}
	\mathcal{S}_{m,n}^\sigma = \left\{ \psi: \mathbb{R}^m \to \mathbb{R}^n \mid \psi(x) = L''(\sigma(L'(x))) \right\},
\end{equation*}
where $L' \in \mathcal{A}(m, l)$ and $L'' \in \mathcal{A}(l, n)$ are affine mappings.

The trunk network is composed of a single hybrid quantum layer $\mathbf{h}$ which is showed in Figure \ref{fig:hybrid system}:
\begin{equation}\label{a1}
	\mathbf{t}(y) = \mathbf{h}(y) = \phi\bigl(\mathbf{f}(\psi(y))\bigr)\in \mathbb{R}^p,
\end{equation}
where $\phi \in \mathcal{S}_{q,p}^\sigma, \psi \in \mathcal{S}_{m,q}^\sigma$.
 $\mathbf{f}: \mathbb{R}^q \to \mathbb{R}^q$ in equation \eqref{a1} is a Parameterized Quantum Circuit, whose $v$-th component is given by:
\begin{equation}
	f_v(z) = \langle 0 |^{q} U^\dagger(z,\boldsymbol{\theta}) M_v U(z,\boldsymbol{\theta}) | 0 \rangle ^{q} \in \mathbb{R}, 
\end{equation}
where $z \in \mathbb{R}^{q}$ and $v = 1, \ldots, q$. Here, $|0\rangle$ is the initial state of the quantum computer, $U(z,\boldsymbol{\theta})$ is a quantum circuit with trainable parameters $\boldsymbol{\theta}$ and $M_v$ is an observable of the $v$-th qubit.  The overall quantum circuit has the form 
\begin{equation*}
U(z,\boldsymbol{\theta})=W^{L}(\boldsymbol{\theta})\cdots W^{2}(\boldsymbol{\theta})W^{1}(\boldsymbol{\theta})S(z),
\end{equation*}
where $W(\boldsymbol{\theta})$ is circuit block controlled by the
trainable parameters $\boldsymbol{\theta}$, S(z) is encoding circuit block.

\begin{center}
\includegraphics[width=0.6\linewidth]{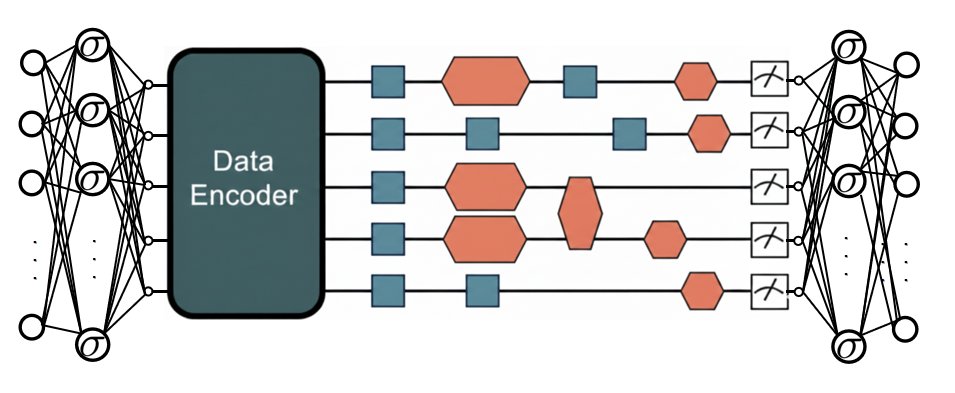}
\captionof{figure}{\label{fig:hybrid system}Architecture of quantum layers.The pre- and post-processing layers of the quantum circuit consist of affine transformations and nonlinear activation functions. An angle encoder is used in the quantum circuit. The circuit employs three distinct structures—Nearest-neighbour, All-to-all, and Circuit-block—corresponding to circuits 2, 6, and 19 in \cite{sim2019expressibility}, respectively. These structures are implemented within the geometrically irregular region following the Data Encoder, as illustrated in Figure \ref{fig:hybrid system}. Measurements in the quantum circuit are performed by calculating the expected value of each qubit under the Pauli-Z operator.}
\end{center}

Furthermore, the performance of a PQC is highly dependent on its internal architectural design\cite{hubregtsen2021evaluation, sim2019expressibility}; studies have found that circuits employing controlled-X rotation (CRX) gates generally exhibit stronger expressibility than those using controlled-Z rotation (CRZ) gates because the commutativity between CRZ gates limits the dimensionality of the state space exploration. Additionally, while an all-to-all topology provides the best expressibility and entangling capability, it incurs high hardware costs, whereas a circuit-block structure offers a better balance between performance and cost. Therefore, this paper employs circuits with various structures—including all-to-all, circuit-block, nearest-neighbor, and quantum optimization algorithm-based designs—and combines them with operator methods to investigate different problems. The numbers of parameters and gates of different circuits are shown in Table \ref{tab:para}. For specific details on the circuit structure, please refer to \cite{sim2019expressibility}.

\begin{center}
\begin{tabular}{llll}
\hline
Ansatz          & Circuit 2 (Nearest-neighbour) & Circuit 6 (All-to-all) & Circuit 19 (Circuit-block) \\ \hline
Parameters      & 4nL                           & n(n+3)L                & 3nL                         \\ \hline
Two-qubit gates & nL                            & n(n-1)L                & nL                          \\ \hline
\end{tabular}
\captionof{table}{Table of the number of parameters and gates.}
\label{tab:para}
\end{center}


The branch network utilizes an architecture consisting of $r$ stacked sub-networks followed by a cross-subnet attention layer to capture cross-subnet relations among tokens from different quantum sub-networks. The sampled input $u_{s} = (u(x_1), \ldots, u(x_i)) \in \mathbb{R}^d$ from $d$ sensor locations is partitioned into $r$ sub-vectors $u^{(1)}_{s}, \ldots, u^{(r)}_{s}$ according to the sensor distribution, where the $j$-th sub-vector $u^{(j)}_{s} \in \mathbb{R}^{d_j}$ satisfies $\sum_{j=1}^r d_j = d$. For simplicity, all sub-vectors and the outputs of sub-networks have the same dimension, i.e.,
\begin{equation*}
		d_1 = d_2 = \cdots = d_r = \frac{d}{r},\quad
		p_1 = p_2 = \cdots = p_r = \mathbf{c} = \frac{p}{r}.
\end{equation*}
Similar to Equation \eqref{a1}, the output of the $i$-th sub-network in branch network is as follows:
\begin{equation}
	\mathbf{h}^{(i)}(u_s^{(i)}) = \left(\phi^{(i)}\bigl(\mathbf{f}^{(i)}(\psi^{(i)}(u_s^{(i)}))\bigr)\right)\in \mathbb{R}^{\mathbf{c}}.
\end{equation}
Rewrite the outputs of r sub-networks into: 
\begin{equation}
\mathbf{H} = [ \mathbf{h}^{(1)}(u_s^{(1)}), \mathbf{h}^{(2)}(u_s^{(2)}), \dots, \mathbf{h}^{(r)}(u_s^{(r)}) ]^{T}.
\end{equation}
Define the global average pooling operator $\mathcal{G} : \mathbb{R}^{r \times \mathbf{c}} \rightarrow \mathbb{R}^{r}$:
\begin{equation}
	\mathbf{z} = \mathcal{G}(\mathbf{H}) = \frac{1}{\mathbf{c}} \mathbf{H} \mathbf{1}_{\mathbf{c}},
\end{equation}
where $\mathbf{1}_{\mathbf{c}} = (1, 1, \dots, 1)^{T} \in \mathbb{R}^{\mathbf{c}}$. This produces a vector $\mathbf{z} \in \mathbb{R}^r$ where each element represents the average activation of the corresponding subnet.

To determine the coverage of the local cross-subnet interaction, define the nonlinear mapping $\psi_k : \mathbb{N} \rightarrow \mathbb{N}_{\mathrm{odd}}$ \cite{2020Wang}. The kernel size $k$ of the 1D convolution is adaptively selected based on the number of subnets $r$:
\begin{equation*}
	k = \psi_k(r) = \left| \frac{\log_2(r)}{\gamma} + \frac{b}{\gamma} \right|_{\text{odd}},
\end{equation*}
where $\gamma, b$ are hyperparameters for the mapping , and $|\cdot|_{\text{odd}}$ denotes the nearest odd number. 


To capture local cross-subnet interaction, a Toeplitz banded matrix $W_k \in \mathbb{R}^{r \times r}$ is employed whose elements are
\begin{equation*}
(W_k)_{i,j} = 
\begin{cases} 
    w_{i-j}, & \text{if } |i - j| < k/2 \\ 
    0, & \text{otherwise} 
\end{cases}
\end{equation*}
Define the subnet attention weight generation function $\mathcal{A} :
	\boldsymbol{\omega} = \mathcal{A}(\mathbf{z}) = \sigma_0(W_k \mathbf{z})
, $
where $\sigma_0(x) = \frac{1}{1+e^{-x}}$. Each element $\omega_i$ represents the learned importance weight for subnet $i$, influenced by the activations of its neighboring subnets through the local 1D convolution.

The attention-modulated subnet outputs are obtained by element-wise scaling:
\begin{equation*}
	\tilde{\mathbf{H}}=\text{diag}(\boldsymbol{\omega})\mathbf{H}:=[ \tilde{\mathbf{h}}^{(1)}(u_s^{(1)}), \tilde{\mathbf{h}}^{(2)}(u_s^{(2)}), \dots, \tilde{\mathbf{h}}^{(r)}(u_s^{(r)}) ]^{T}.
\end{equation*}
 This operation uniformly scales all $\mathbf{c}$ features of each subnet $i$ by its corresponding attention weight $\omega_i$.

The final output of the branch network, where $p = \mathbf{c} \times r$, is given by
\begin{equation}
	\mathbf{b}(u_{s})= \begin{bmatrix}
		\tilde{\mathbf{h}}^{(1)}(u_s^{(1)}) \\
		\tilde{\mathbf{h}}^{(2)}(u_s^{(2)}) \\
		\vdots \\
		\tilde{\mathbf{h}}^{(r)}(u_s^{(r)})
	\end{bmatrix} \in \mathbb{R}^p.
\end{equation}

\textbf{Remark:} Unlike the ECA method that aggregates information across spatial tokens, our approach targets \emph{cross-subnet interaction}. Therefore, we perform global average pooling along the \emph{feature dimension} rather than the subnet dimension. The proposed subnet attention mechanism has only $k$ learnable parameters, compared to hundreds of parameters in multi-head attention variants, while maintaining $O(r)$ computational complexity for processing $r$ subnets.

The output is computed via the inner product of the $p$-dimensional outputs from the trunk and branch networks:
\begin{equation}
	G(u)(y) = \langle \mathbf{b}(u_{s}), \mathbf{t}(y) \rangle := \sum_{k=1}^{p} b_k(u_{s}) \cdot t_k(y) \in \mathbb{R},
\end{equation}
where $\mathbf{b}(u_s), \mathbf{t}(y) \in \mathbb{R}^p$, and $b_k, t_k$ represent their respective $k$-th components.

\begin{center}
\hspace{-2.9cm}
\begin{minipage}{0.45\textwidth}
\centering
\includegraphics[width=1.0\linewidth, height=5cm]{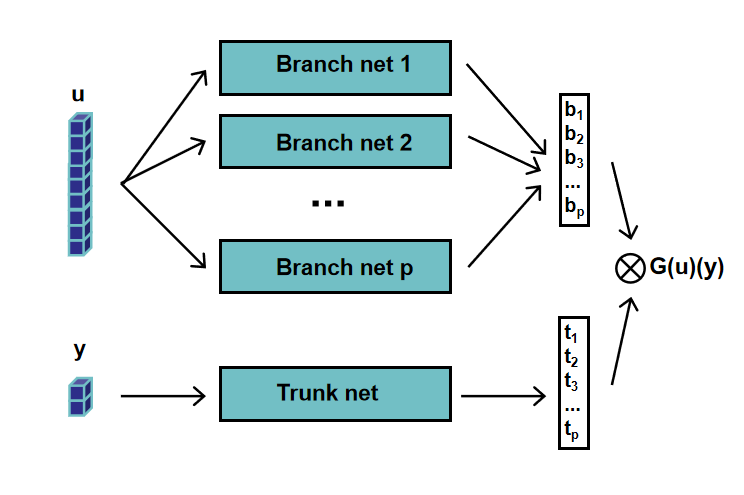}
\captionof{figure}{\label{fig:deeponet}Stacked DeepONet structure}
\end{minipage}
\hspace{0.01\textwidth}
\begin{minipage}{0.45\textwidth}
\centering
\includegraphics[width=1.3\linewidth, height=5cm]{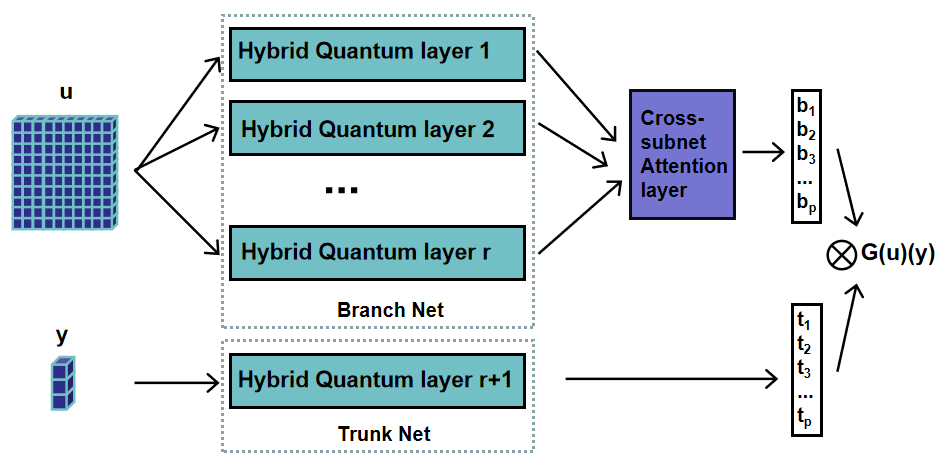}
\captionof{figure}{\label{fig:liucheng}Architecture of Quantum AS-DeepOnet. }
\end{minipage}
\end{center}
\textbf{Remark:} Different from the stacked DeepONet structure in \cite{2021lulu}, where the number of sub-networks in the branch net equals the output dimension, Quantum AS-DeepOnet uses several sub-networks to share the input information and then aggregates the outputs of these sub-networks through a cross-subnet attention layer. This design allows for a more flexible and efficient representation of the input function, as it can capture complex interactions between different parts of the input through the attention mechanism, while also reducing the overall number of parameters.

\section{Numerical results and discussion}\label{numerical experiments}

In this section, we present numerical results to demonstrate the performance of the proposed Quantum AS-DeepOnet. We consider two benchmark problems: the 2D advection equation and the 2D Burgers' equation.

\begin{example}
We consider the two-dimensional linear advection equation defined on the spatial-temporal domain $\Omega = [0,1] \times [0,1] \times [0,1]$:

\begin{equation*}
	\frac{\partial u}{\partial t} + v_x \frac{\partial u}{\partial x} + v_y \frac{\partial u}{\partial y} = 0, \quad (x,y,t) \in \Omega,
\end{equation*}

\noindent subject to periodic boundary conditions in both spatial directions:
\begin{equation*}
		u(x, 0, t) = u(x, 1, t),\quad u(0, y, t) = u(1, y, t),\quad x \in [0,1],~ y \in [0,1],~ t \in [0,1].
\end{equation*}

Here, $u(x,y,t)$ represents the advected quantity (e.g., concentration, temperature), and $(v_x, v_y) = (1.0, 0.5)$ is the constant advection velocity vector.

The initial condition $u_0(x,y)$ is generated as a two-dimensional Gaussian random field (GRF) to represent various spatial distributions. The GRF is constructed using the Fourier transform method with a Gaussian power spectral density:
\begin{equation}\label{gauss}
	S(k_x, k_y) = A \exp\left(-\frac{(k_x \ell_x)^2 + (k_y \ell_y)^2}{2}\right),
\end{equation}
where $A$ is the amplitude, and $(\ell_x, \ell_y)$ are the correlation length scales in the $x$ and $y$ directions, respectively. Random phases are uniformly distributed in $[0, 2\pi]$, and Gaussian smoothing with standard deviation is applied to ensure spatial continuity. The resulting random field is normalized to a desired range to ensure numerical stability during time integration.
\end{example}




\begin{example}
We consider the two-dimensional nonlinear Burgers' equation defined on the spatial-temporal domain $\Omega = [0,1] \times [0,1] \times [0,1]$:
\begin{equation*}
	\frac{\partial u}{\partial t} + u\frac{\partial u}{\partial x} + u\frac{\partial u}{\partial y} = \nu\left(\frac{\partial^2 u}{\partial x^2} + \frac{\partial^2 u}{\partial y^2}\right), \quad (x,y,t) \in \Omega,
\end{equation*}

\noindent subject to periodic boundary conditions in both spatial directions. Here, $u(x,y,t)$ represents the velocity field (or concentration in other applications), $\nu$ is the kinematic viscosity coefficient, and the equation captures both the nonlinear advection (convection) term and the viscous (diffusive) effects. 

The initial condition $u_0(x,y)$ is generated as a two-dimensional Gaussian random field (GRF) to represent various spatial distributions. The GRF is constructed using the Fourier transform method with a Gaussian power spectral density as described in equation \eqref{gauss}, where the correlation length scales are set to $(\ell_x, \ell_y) = (0.4, 0.4)$ to create more localized structures in the initial condition. 
\end{example}
During training, we uniformly employ the Adam optimizer with an initial learning rate of 0.002 and a three-stage learning rate scheduler. The training runs for 60,000 epochs, with model performance evaluated every 500 epochs. The training set contains 10,000 samples, and the test set also contains 10,000 samples. We compare the performance of different quantum circuit structures (nearest-neighbor, all-to-all, and circuit-block) as well as two classical models. Specifically, for the Advection equation, we use a 10-qubit circuit, and for Burgers' equation, we use a 12-qubit circuit. Both quantum circuits have a depth of 2, with pre- and post-processing layer widths of 50. Classical Model 1 and Classical Model 2 employ the classical DeepONet method with a width of 50 and different depths. Specifically, Classical Model 1 has a depth of 4, and Classical Model 2 has a depth of 2. Table \ref{tab:performance_comparison} summarizes the number of parameters, relative L2 error, and final loss value for each method. Figures \ref{fig:compare1} and \ref{fig:compare2} show the comparison between the prediction results and true solutions for different methods on the 2D advection equation and 2D Burgers' equation. 

The results demonstrate that Quantum AS-DeepOnet achieves comparable or even superior performance compared to classical models while maintaining a lower parameter count, particularly showing stronger expressiveness and generalization ability on Burgers' equation. This indicates that the choice of quantum circuit structure has a significant impact on model performance and validates the effectiveness of our proposed Quantum AS-DeepOnet architecture for solving complex PDE problems. Current results are obtained on the PennyLane simulator. Although this model reduces computational complexity and achieves acceleration through Parameterized Quantum Circuits, the training speed of the Quantum AS-DeepOnet is slower than the classical DeepOnet. This is due to time-consuming bottlenecks such as classical-to-quantum data conversion and the inherent limitations of current quantum simulators. 
\begin{center}
\resizebox{\textwidth}{!}{%
\begin{tabular}{lllllll}
\hline
                           & Ansatz                 & Circuit-block & Nearest-neighbour & All-to-all & Classical Model 1 & Classical Model 2 \\ \hline
\multirow{3}{*}{Advection} & Parameters             & 13292         & 13392             & 14292      & 24251             & 14051             \\ \cline{2-7} 
                           & Relative L2 error (\%)  & 3.38e-02      & 7.60e-02          & 3.51e-02   & 3.42e-02          & 9.08e-02          \\ \cline{2-7} 
                           & Last loss              & 3.09e-04      & 1.52e-03          & 3.60e-04   & 3.17e-04          & 2.17e-03          \\ \hline
\multirow{3}{*}{Burgers}   & Parameters             & 14342         & 14442             & 15342      & 24251             & 14051             \\ \cline{2-7} 
                           & Relative L2 error (\%) & 4.36e-02      & 5.45e-02          & 3.31e-02   & 3.23e-02          & 8.44e-02          \\ \cline{2-7} 
                           & Last loss              & 5.89e-06      & 8.76e-06          & 3.22e-06   & 3.05e-04          & 2.02e-05          \\ \hline
\end{tabular}
}%
\captionof{table}{Comparison of different quantum circuit ansätze for 2D Advection and Burgers' equations.}
\label{tab:performance_comparison}
\end{center}

\begin{center}
\begin{minipage}{0.48\textwidth}
\centering
\includegraphics[width=\textwidth]{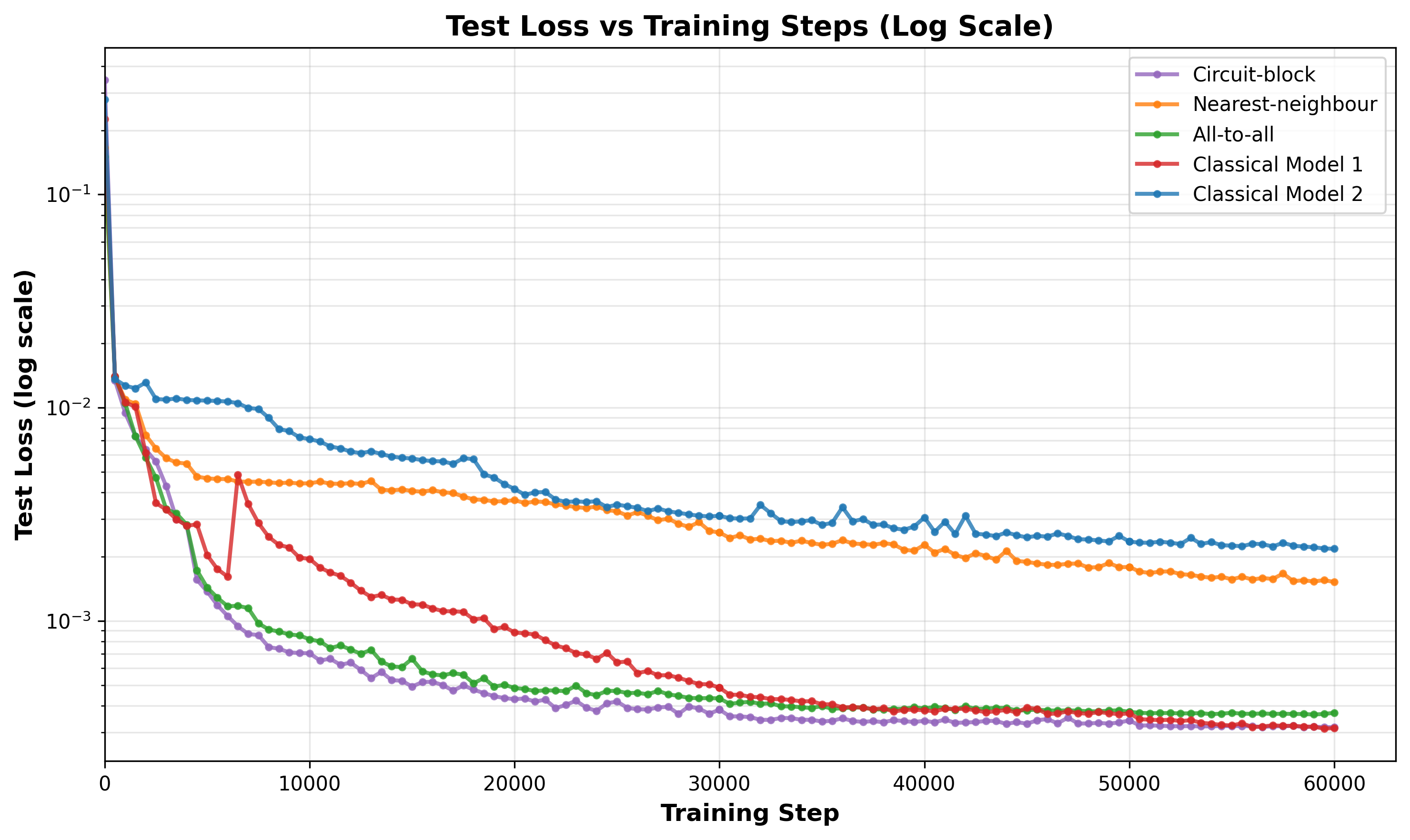}
\captionof{figure}{Comparison results for 2D advection equation.}
\label{fig:compare1}
\end{minipage}
\hspace{0.01\textwidth}
\begin{minipage}{0.48\textwidth}
\centering
\includegraphics[width=\textwidth]{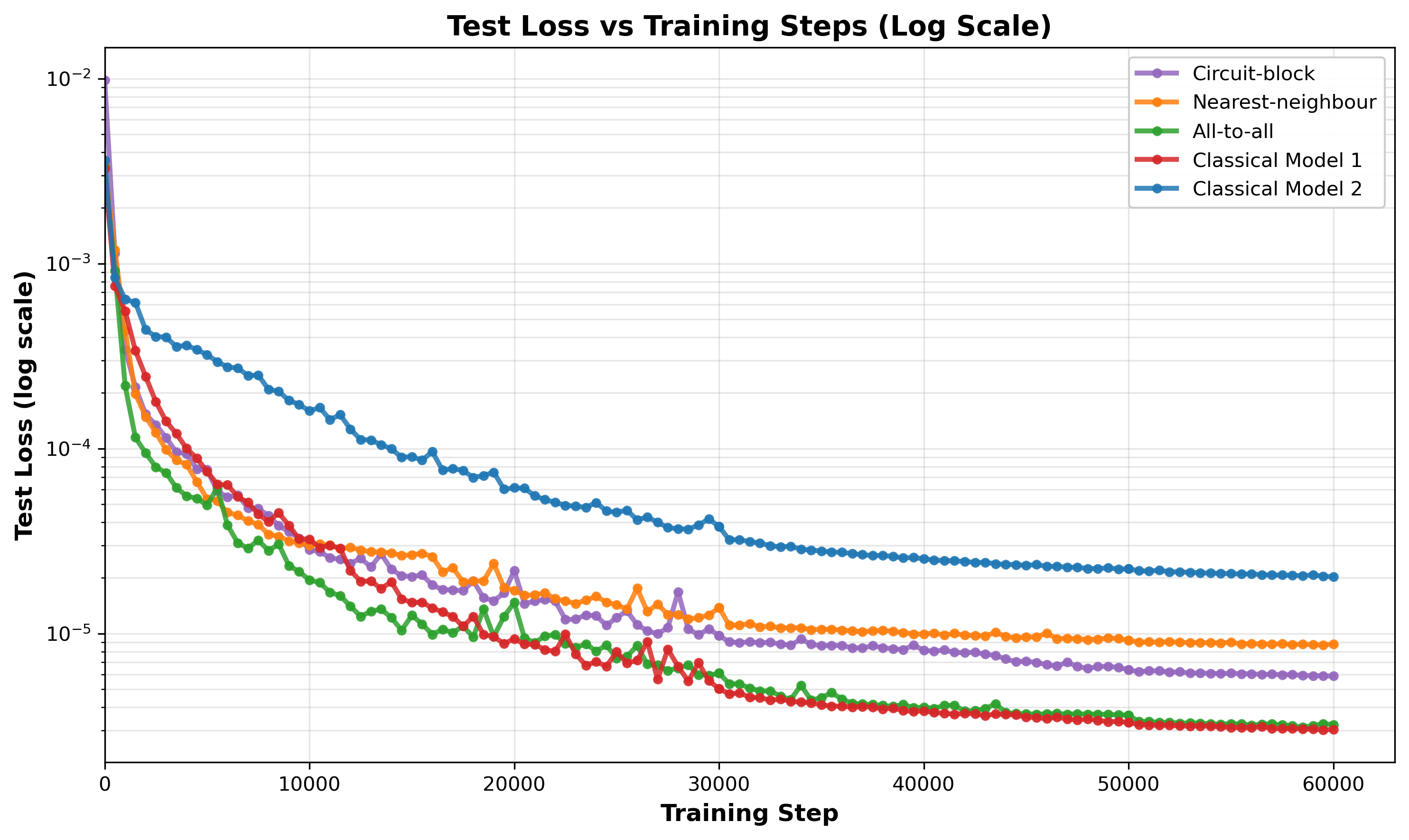}
\captionof{figure}{Comparison results for 2D Burgers' equation.}
\label{fig:compare2}
\end{minipage}
\end{center}


\section{Conclusion}

In conclusion, this work presents Quantum AS-DeepONet, a novel hybrid quantum operator network for solving 2D evolution equations. The proposed model achieves accuracy and convergence comparable to classical DeepONet while using significantly fewer parameters. This hybrid approach offers a scalable method for operator learning, though further research is needed to enhance its resilience against quantum hardware noise. Future work includes deploying the model on quantum hardware to validate its performance and practical applicability.

\section*{Acknowledgments}
	This work was supported by the National Science Foundation of China (12372196).


\bibliographystyle{unsrt}

\bibliography{cas-refs}

\begin{thebibliography}{10}

\bibitem{2019Raissi}
M.~Raissi, P.~Perdikaris, and G.~E. Karniadakis.
\newblock Physics-informed neural networks: A deep learning framework for
  solving forward and inverse problems involving nonlinear partial differential
  equations.
\newblock {\em Journal of Computational Physics}, 378:686--707, 2019.

\bibitem{2020Pang}
G.~Pang, M.~D'Elia, M.~Parks, and G.~E. Karniadakis.
\newblock npinns: Nonlocal physics-informed neural networks for a parametrized
  nonlocal universal laplacian operator. algorithms and applications.
\newblock {\em Journal of Computational Physics}, 422:109760, 2020.

\bibitem{2021lulu}
L.~Lu, P.~Jin, G.~Pang, et~al.
\newblock Learning nonlinear operators via deeponet based on the universal
  approximation theorem of operators.
\newblock {\em Nature Machine Intelligence}, 3(3):218--229, 2021.

\bibitem{2019Schuld}
M.~Schuld and N.~Killoran.
\newblock Quantum machine learning in feature hilbert spaces.
\newblock {\em Physical Review Letters}, 122(4):040504, 2019.

\bibitem{2019V}
V.~Havlíček, A.~D. Córcoles, K.~Temme, et~al.
\newblock Supervised learning with quantum-enhanced feature spaces.
\newblock {\em Nature}, 567(7747):209--212, 2019.

\bibitem{2021Kwak}
Y.~Kwak, W.~J. Yun, S.~Jung, et~al.
\newblock Introduction to quantum reinforcement learning: Theory and
  pennylane-based implementation.
\newblock In {\em 2021 International Conference on Information and
  Communication Technology Convergence (ICTC)}, pages 416--420. IEEE, 2021.

\bibitem{2018Amin}
M.~H. Amin, E.~Andriyash, J.~Rolfe, et~al.
\newblock Quantum boltzmann machine.
\newblock {\em Physical Review X}, 8(2):021050, 2018.

\bibitem{2021Xin}
T.~Xin, L.~Che, C.~Xi, et~al.
\newblock Experimental quantum principal component analysis via parametrized
  quantum circuits.
\newblock {\em Physical Review Letters}, 126(11):110502, 2021.

\bibitem{2024li}
J.~Li, Y.~Li, J.~Song, et~al.
\newblock Quantum support vector machine for classifying noisy data.
\newblock {\em IEEE Transactions on Computers}, 73(9):2233--2247, 2024.

\bibitem{2025Farea}
A.~Farea, S.~Khan, and M.~Serdar~Celebi.
\newblock Qcpinn: quantum-classical physics-informed neural networks for
  solving pdes.
\newblock {\em Machine Learning: Science and Technology}, 6(4):045053, 2025.

\bibitem{2024Baker}
J.~S. Baker, G.~Park, K.~Yu, et~al.
\newblock Parallel hybrid quantum-classical machine learning for kernelized
  time-series classification.
\newblock {\em Quantum Machine Intelligence}, 6(1):18, 2024.

\bibitem{2019Benedetti}
M.~Benedetti, E.~Lloyd, S.~Sack, and M.~Fiorentini.
\newblock Parameterized quantum circuits as machine learning models.
\newblock {\em Quantum Science and Technology}, 4(4):043001, 2019.

\bibitem{2025Wang}
R.~Wang, Z.~Xia, G.~Yan, et~al.
\newblock Quanonet: Quantum neural operator with application to differential
  equation.
\newblock In {\em Forty-second International Conference on Machine Learning},
  2025.

\bibitem{2025Xiao}
P.~Xiao, M.~Zheng, A.~Jiao, et~al.
\newblock Quantum deeponet: Neural operators accelerated by quantum computing.
\newblock {\em Quantum}, 9:1761, 2025.

\bibitem{sim2019expressibility}
Sukin Sim, Peter~D. Johnson, and Alán Aspuru-Guzik.
\newblock Expressibility and entangling capability of parameterized quantum
  circuits for hybrid quantum-classical algorithms.
\newblock {\em Advanced Quantum Technologies}, 2(12):1900070, 2019.

\bibitem{hubregtsen2021evaluation}
Thomas Hubregtsen, Josef Pichlmeier, Patrick Stecher, and Koen Bertels.
\newblock Evaluation of parameterized quantum circuits: on the relation between
  classification accuracy, expressibility, and entangling capability.
\newblock {\em Quantum Machine Intelligence}, 3(1):9, 2021.

\bibitem{2020Wang}
Q.~Wang, B.~Wu, P.~Zhu, et~al.
\newblock Eca-net: Efficient channel attention for deep convolutional neural
  networks.
\newblock In {\em Proceedings of the IEEE/CVF Conference on Computer Vision and
  Pattern Recognition}, pages 11534--11542, 2020.

\end{thebibliography}



\end{document}